\documentclass{iaus}
\usepackage{graphicx}
\usepackage{natbib}
\usepackage{url}
\usepackage{hyperref}

\usepackage{multicol}

\def\aj{AJ}%
\def\apj{ApJ}%
\def\apjl{ApJ}%
\def\apss{Ap\&SS}%
\def\aap{A\&A}%
\def\mnras{MNRAS}%
\def\nat{Nature}%
\def\hi{H{\sc I}}

\title[SED of Edge-on Spirals]{New HErschel Multi-wavelength Extragalactic Survey of Edge-on Spirals (NHEMESES)}
\author[B.W. Holwerda et al.]   
{B.W. Holwerda$^1$, S. Bianchi$^2$, M. Baes$^3$, R.S. de Jong$^4$, J.J. Dalcanton$^5$, D. Radburn-Smith$^5$, K. Gordon$^6$ \and M. Xilouris$^7$}

\affiliation{$^1$ European Space Agency (ESTEC),
Keplerlaan 1, 2200 AV Noordwijk, The Netherlands\\

email: {\tt \href{mailto:benne.holwerda@esa.int}{benne.holwerda@esa.int}} \\[\affilskip]
$^2$ INAF-Arcetri Astrophysical Observatory, Florence,
$^3$ University of Gent,
$^4$ Astronomisch Insit\"{u}t Potsdam,
$^5$ University of Washington,
$^6$ Space Telescope Science Institute,
$^7$ National Observatory of Athens
}

\pubyear{2011} 
\volume{284}  
\pagerange{1--12}
\jname{The Spectral Energy Distribution of Galaxies}
\editors{R.J. Tuffs \&  C.C.Popescu, eds.}
\begin{document}

\maketitle

\begin{abstract}

Edge-on spiral galaxies offer a unique perspective on the vertical structure of spiral disks, both stars and the iconic dark dustlanes. The thickness of these dustlanes can now be resolved for the first time with {\em Herschel} in far-infrared and sub-mm emission. We present NHEMESES, an ongoing project that targets 12 edge-on spiral galaxies with the PACS and SPIRE instruments on {\em Herschel}. These vertically resolved observations of edge-on spirals will impact on several current topics.

First and foremost, these {\em Herschel} observations will settle whether or not there is a phase change in the vertical structure of the ISM with disk mass. Previously, a dramatic change in dustlane morphology was observed as in massive disks the dust collapses into a thin lane. If this is the case, the vertical balance between turbulence and gravity dictates the ISM structure and consequently star-formation and related phenomena (spiral arms, bars etc.). We specifically target lower mass nearby edge-ons to complement existing {\em Herschel} observations of high-mass edge-on spirals (the HEROES project).

Secondly, the combined data-set, together with existing {\em Spitzer} observations, will drive a new generation of spiral disk Spectral Energy Distribution models. 
These model how dust reprocesses starlight to thermal emission but the dust geometry remains the critical unknown. 

And thirdly, the observations will provide an accurate and unbiased census of the cold dusty structures occasionally seen extending out of the plane of the disk, when backlit by the stellar disk. To illustrate the NHEMESES project, we present early results on NGC 4244 and NGC 891, two well studies examples of a low and high-mass edge-on spiral.
\end{abstract}


Edge-on spiral galaxies offer a unique perspective on spiral disks. An observer can explore the vertical structure of both stars and ISM using different wavelengths. Dust is mechanically linked to the cold ISM \citep{Allen86, Weingartner01b}. \cite{Dalcanton04} used the appearance of dust lanes as a probe of vertical stability of spiral disks. They found that in massive spiral disks, the ISM collapses into a thin dustlane, while the less massive disks show more flocculant dust morphology. In addition, observations of edge-ons often reveal dark structures extending out of the plane of massive disks \citep[e.g.,][]{Howk99a}, but the quantity of extra-planar dust has been impossible to constrain from extinction measures.

A comprehensive approach is to model multi-wavelength images of edge-on spirals with a bulge, a stellar and a dust disk to constrain stellar and ISM structure \citep[e.g.,][]{Xilouris99, Bianchi08, Baes10a, Popescu11}. However, until now, these models are often degenerate in the vertical distributions of stellar light and ISM due to lack of resolution and wavelength coverage. With the advent of {\em Herschel}, the vertical structure of some nearby edge-on disks can be now resolved. Our {\em Herschel} program NHEMESES\footnote{{\bf N}ew {\bf HE}rshel {\bf M}ulti-wavelength {\bf E}xtragalactic {\bf S}urvey of {\bf E}dge-on {\bf S}pirals, a 10.3 hour, OT1, priority 2 program to supplement the GT program on large disks,
{\bf HER}shel {\bf O}bservations {\bf E}dge-on {\bf S}pirals, PI M. Baes.} (PI B.W. Holwerda) aims to (1) measure the vertical scale of dust in spiral disks (an unknown in the above models), (2) the extra-planer dust component, and (3) whether or not the strong break seen in dust lanes by \cite{Dalcanton04} is indeed a real phase-change in the cold ISM with disk mass.


\begin{figure}
  \begin{center}
    \begin{minipage}[l]{0.5\linewidth}
	\includegraphics[width=\textwidth]{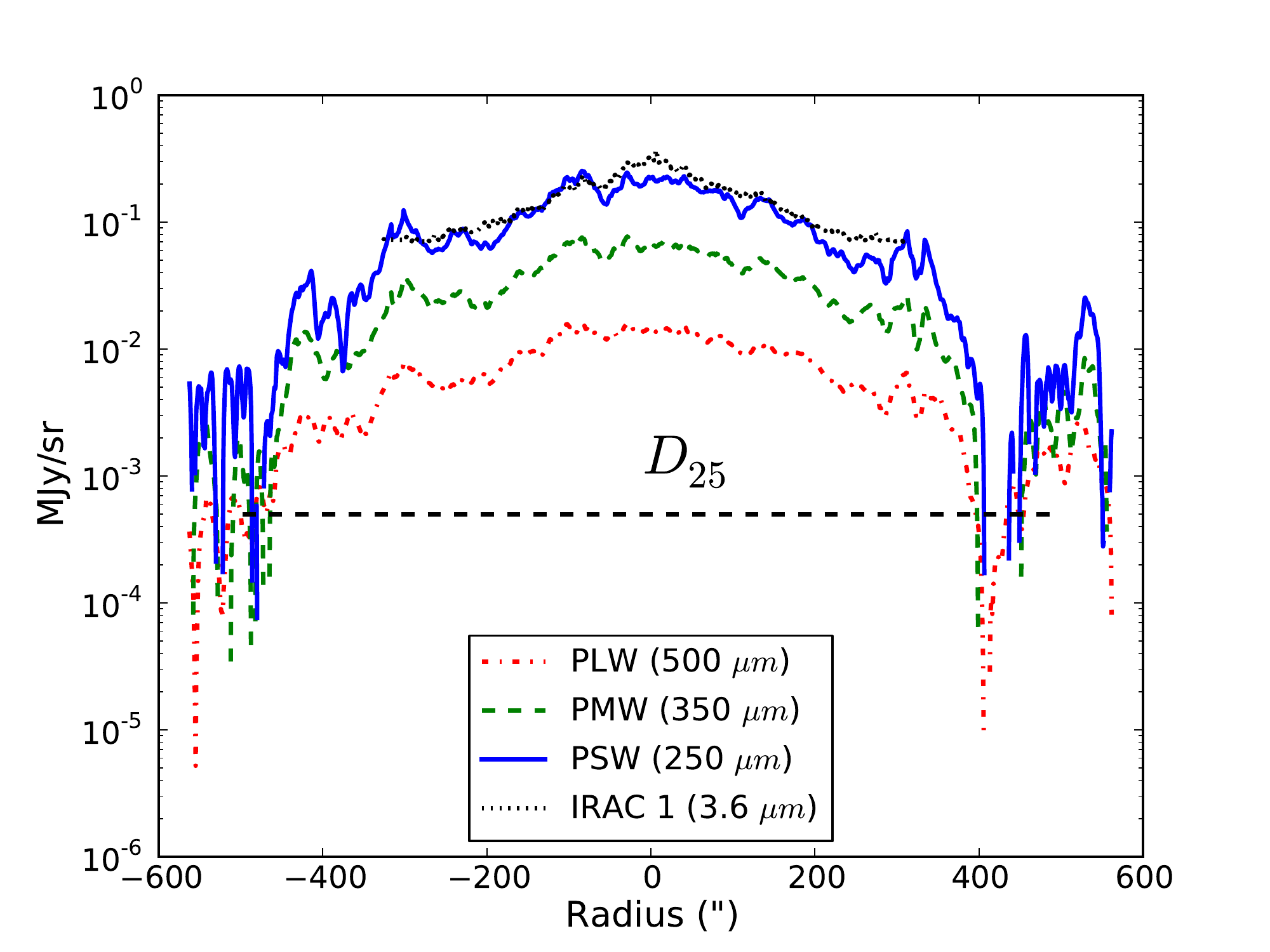} 	
 	\caption{\label{f:bwh:f1} The radial profiles of NGC4244; {\em Spitzer} 3.6 (black), SPIRE 250 (red), 350 (green) and 500 (blue) $\mu$m.} 
    \end{minipage}\hfill
    \begin{minipage}[r]{0.5\linewidth}
	\includegraphics[width=\textwidth]{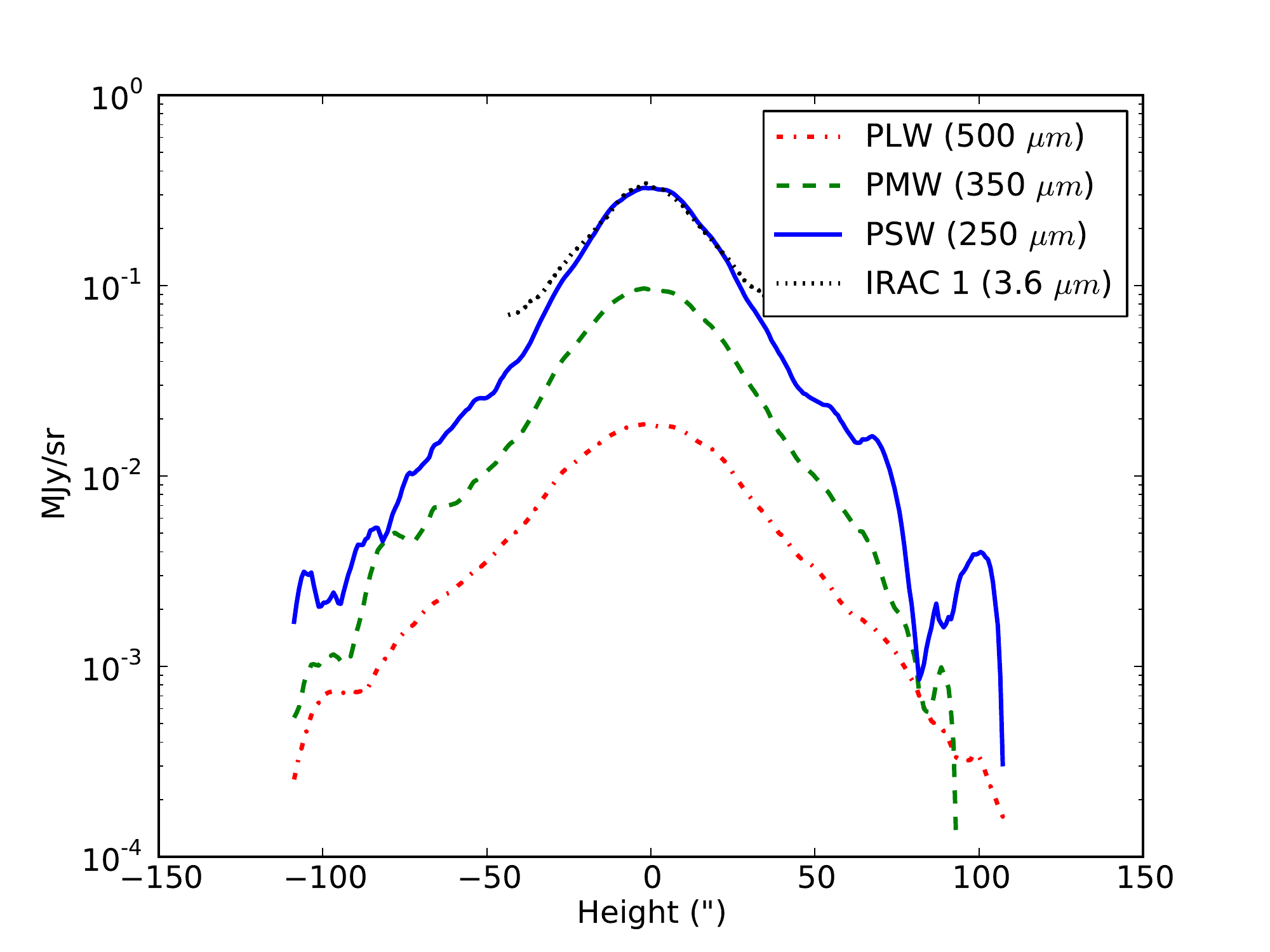} 	
	\caption{\label{f:bwh:f2} The vertical profile of NGC4244, same legend as Fig. \ref{f:bwh:f1}. The SPIRE profile stays within the 3.6 $\mu$m height.} 
      \end{minipage}
  \end{center}
\end{figure}

We show some first NHEMESES results with the SPIRE morphology of NGC 4244, the prototypical disk-dominated low-mass spiral galaxy ($\rm v_{rot} = 95 ~ km s^{-1}$). 

\begin{figure}
  \begin{center}
    \begin{minipage}[h]{0.55\linewidth}
 	\includegraphics[width=\textwidth]{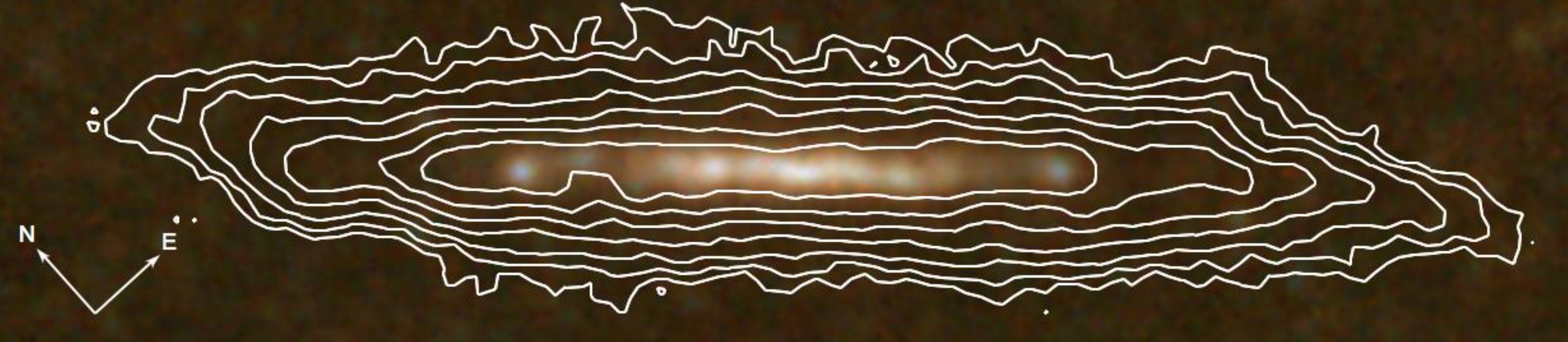}
 	\caption{\label{f:bwh:f3} A three-color image of the SPIRE observations of NGC 4244 with 250 (red), 350 (green) and 500 (blue) $\mu$m and the  \hi\ contours from \cite{Zschaechner11a} (contours at $6.4 \times 10^{19}$ cm$^{-2}$ increasing by factors of 2). The SPIRE flux is contained mostly within the $8.1 \times 10^{19}$ cm$^{-2}$ contour.}
 	\caption{   \label{f:bwh:f4} The Spectral Energy Distribution model from \cite{MacLachlan11} (see also MacLachlan et al.({\em this volume}) and the SPIRE fluxes for NGC 4244. The SED under-predicts the sub-mm fluxes by only a factor 2.}        
    \end{minipage}\hfill
    \begin{minipage}[h]{0.44\linewidth}
 	\includegraphics[width=\textwidth]{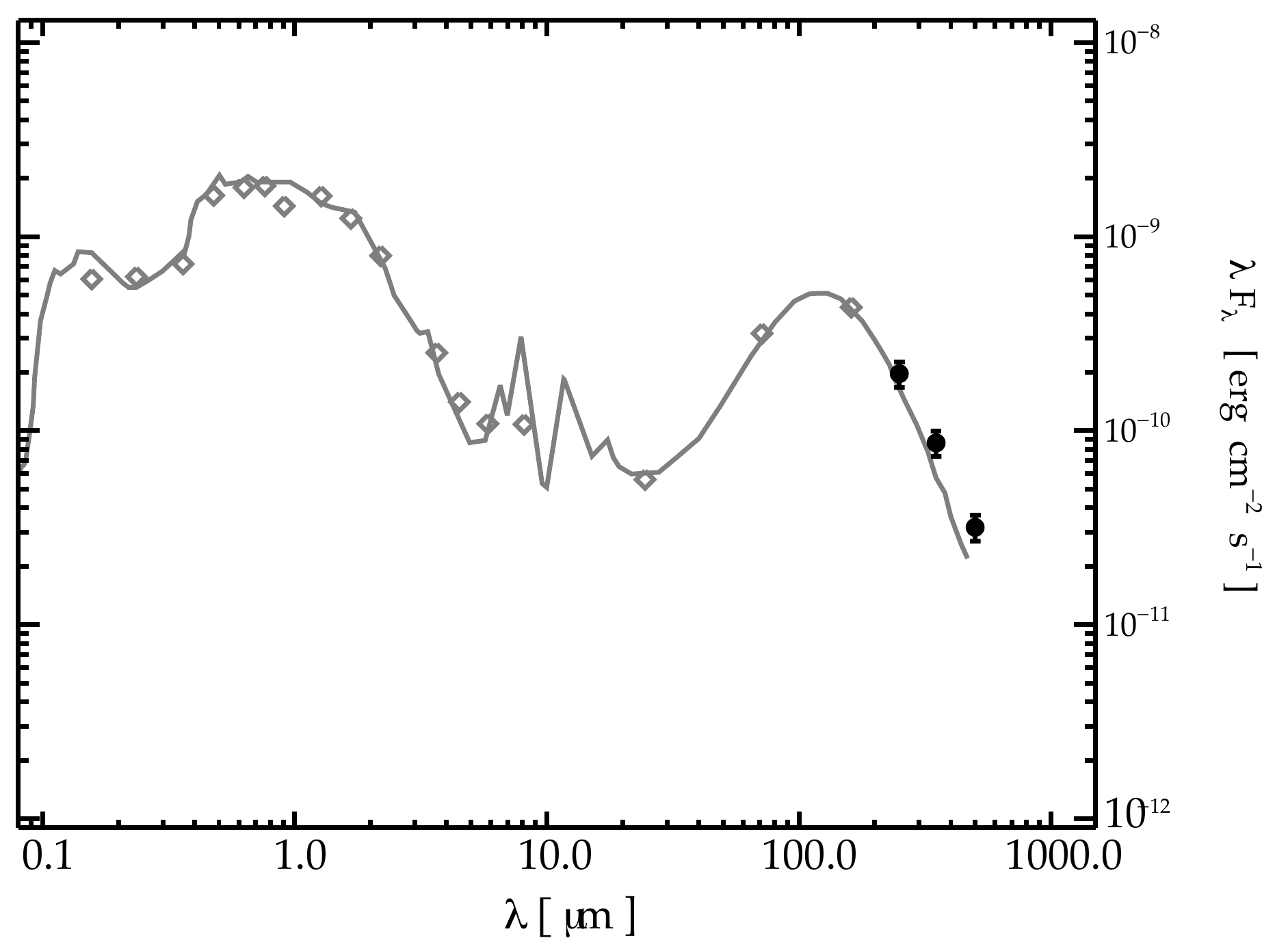} 
    \end{minipage}
  \end{center}
\end{figure}

The radial profile (Fig. \ref{f:bwh:f1}) appears to truncate beyond a peak in flux from a small star-formation region on both sides of the disks (see also Fig. \ref{f:bwh:f3}). The radius of truncation is similar to those initially found by \cite{vdKruit81a} and more recently by \cite{de-Jong07} for different stellar populations.

The vertical sub-mm profiles (Fig. \ref{f:bwh:f2}) are similar in width as the {\em Spitzer} 3.6 $\mu$m emission, a good tracer of the stellar mass (e.g., Meidt et al., {\em this volume}). \cite{Comeron11a} find evidence for a thick and thin stellar disk in this galaxy and the dusty ISM appears associated with the inner (thin) disk. In contrast, a second, thicker vertical component in NGC 891 have been reported \citep[][Seon et al., this volume]{Kamphuis07}. 

Fig. \ref{f:bwh:f3} shows the color image based on the 250, 350 and 500 $\mu$m SPIRE images with the \hi \ contours from \cite{Zschaechner11a, Heald11}. The dust emission is restricted to the highest \hi \ contour; it is a single, concentrated disk. This is in contrast to the more massive NGC 891, where \cite{Popescu03, Bianchi11} find evidence for dust throughout the \hi \ disk and extended envelope \citep{Oosterloo07}. 

The Spectral Energy Distribution (SED) model by \cite{MacLachlan11}, in Fig \ref{f:bwh:f4}, finds an optically thin disk with s similar scale-height for both dust and stars \citep[as did][based on stellar populations]{Seth05a}. They report a dust mass of $ 2.38 \times 10^6 M_\odot $ and a dust scale-length 1.8 times its stellar scale length. 
We find that their predicted sub-mm fluxes are lower than what we observe, possibly because the dusty ISM is much clumpier than the smooth diffuse ISM they assume for the SED model. However, we find that their high value of the scale-length of the dusty disk strokes with the radial profile of sub-mm emission we observe (Fig. \ref{f:bwh:f1}).

We find that dust in NGC 4244 is distributed vertically throughout the (thin) stellar disk, in accordance with the prediction from \cite{Dalcanton04}, but that, unlike the much more massive NGC 891, there is little evidence for a second dust component associated with the outer \hi \ envelope. SED models for this galaxy are converging to a solution which includes a large scales (-height and -length) for the dust compared to the stellar ones but distributed almost exclusively in clumps.
This different distribution of the dusty geometry for lower-mass spiral disks may help explain the colder dust temperature and higher relative dust mass observed by H-ATLAS \citep[][Dunne et al. {\em this volume}]{Dunne11}. 

The NHEMESES project aims to observe 12 low-mass spirals with PACS and SPIRE on board {\em Herschel}.
Combined with existing {\em Herschel} observations of massive edge-on spiral galaxies, we aim to (1) ascertain if the ISM indeed goes through a phase-change at 120 km/s, (2) take a census of dusty outflows in nearby, quiescent, spiral disks and (3) provide a suite of multi-wavelength data (in combination with SDSS and {\em Spitzer} observations) to serve as a benchmark for SED models of edge-on spirals galaxies.

\begin{multicols}{2}
\raggedcolumns
\addtocounter{unbalance}{1}

\end{multicols}

\end{document}